\documentclass{ws-procs975x65}

\usepackage{graphicx}
\usepackage{amsmath}
\usepackage{amsfonts}
\usepackage{amssymb}
\usepackage{amsbsy}

\usepackage{colordvi}

\usepackage[active]{srcltx}

\newcommand{\Hil}{\mathcal{H}}

\newcommand{\re}{\mathbb{R}}
\newcommand{\Pl}{{\rm Pl}}
\newcommand{\lPl}{\ell_{\Pl}}

\newcommand{\rd}{{\rm d}}

\newcommand{\id}{\mathbb{I}}

\newcommand{\sgn}{{\rm sgn}}

\begin{document}

\title{\uppercase{Geometric time in quantum cosmology}}

\author{\uppercase{Tomasz Paw{\L}owski}}

\address{Department of Mathematical Methods in Physics, Faculty of Physics,\\
University of Warsaw, Ho\.za 74, 00-682 Warsaw, Poland\\
E-mail: tpawlow@fuw.edu.pl}

\begin{abstract} 
Various choices of the geometry degrees of freedom as the emergent time are tested 
on the model of an isotropic universe with a scalar field of $\phi^2$ potential. 
Potential problems with each choices as well as possible applications in loop quantization
are discussed.
\end{abstract}


\vskip 1.0 cm 
\bodymatter\bigskip

\noindent{{\bf Introduction:}}
One of the main difficulties in quantum gravity/cosmology is the time reparametrization 
invariance, which implies lack of an unambiguous time variable. In consequence,
providing a precise and physically meaningful notion of the system evolution 
--a task particularly crucial in Loop Quantum Cosmology (LQC)-- 
is nontrivial. Usually it is achieved by either deparametrization or the partial 
observable formalism, 
however in order to be practical both techniques require selection of a suitable 
function of the system's degrees of freedom as an internal clock. So far the matter 
degrees of freedom have been chosen for that purpose\cite{aps-imp}. This however 
has made the description dependent on the presence of the particular matter content, 
restricting its applicability. Providing a universal treatment requires using 
the geometry degrees of freedom as a clock.

In the case of isotropic cosmological models there are two obvious choices: volume 
and its canonical momentum (proportional to Hubble parameter), although in LQC 
the application of the former is impared by the bounce phenomenon. Here we test 
the latter choice on the model of a toroidal ($T^3$) FRW universe with massive scalar 
field (the inflaton $\phi^2$ potential) quantized within framework of geometrodynamics 
(Wheeler-DeWitt). 
To define the time evolution we use the deparametrization technique, which poses 
its own challenge as it leads to the (not yet completely understood) $2$nd order 
quantum mechanical formalism with explicit time dependence.
We explore one possible way of defining the suitable formalism using the mapping 
between the ``frozen time'' spaces. The treatment is compared against the textbook 
one, where the scale factor (or volume) plays the role of time. We focus 
on the properties of the ground state needed to tackle the vacuum energy problem
 -- an aspect especially relevant in more realistic (inhomogeneous) cosmological models.

\noindent{{\bf The model:}} The isotropic $T^3$ FRW universe is described by the 
metric
  $g = -N^2\rd t^2 + a^2(t)(\rd\theta^2+\rd\phi^2+\rd\chi^2)$ 
%
where $\theta,\phi,\chi\in[0,1)$, $N$ is the lapse and $a$ is the scale factor. 
Starting from Einstein-Hilbert action for gravity minimally coupled to a massive 
scalar field of mass $m$ (with $\phi^2$ potential) and implementing the canonical 
formalism we arrive to a phase space, which we choose to coordinatize by two pairs 
$(v,b), (\phi,p_{\phi})$, where $v=\alpha^{-1}a^3$, ($\alpha\approx 1.35\lPl^3$), 
$\{v,b\} = 2$, $\phi$ is the scalar field and $p_{\phi}$ its momentum:
$\{\phi,p_{\phi}\}=1$. The dynamics is generated by a Hamiltonian constraint
\begin{equation}\label{eq:constr-class}
  H(v,b,\phi,p_{\phi}) \propto -3\pi G v^2 b^2 + p_{\phi}^2 
  + \alpha^2m^2v^2\phi^2 = 0 .
\end{equation}

\noindent{{\bf Quantization:}} To build the quantum description we follow the 
methods of geometrodynamics, using the elements of a Dirac program. The 
variables $(v,b,\phi,p_{\phi})$ are promoted to operators on the kinematical 
Hilbert space 
$L^2(\re,\rd v)\otimes L^2(\re,\rd\phi)$. 
The quantum counterpart of the Hamiltonian constraint takes the form (with $v=\exp(t)$,
$\hat{v}=e^t\id$ and $\hat{v}\hat{b}=i\partial_t$)
\begin{equation}\label{eq:constr}
  -\partial^2_t \Psi(t,\phi) := (\hat{v}\hat{b}/2)^2 
  = \hat{\Theta}_t \Psi(t,\phi)
  := (12\pi G)^{-1}[\hat{p}_{\phi}^2 + \alpha^2 m^2 e^{2t} \hat{\phi}^2] \Psi(t,\phi) 
\end{equation}
and the physical Hilbert space is composed of the states anihilated 
by it.

\noindent{{\bf Volume deparametrization:}} The Klain-Gordon like form of the constraint 
allows to solve it by the deparametrization (on the quantum level) with respect to $t$. 
The evolution becomes then the mapping between 
the constant $t$ slices of the physical state. However, unlike in \cite{aps-imp} the 
evolution operator $\Theta_t$ generating this mapping is now time dependent. To account 
for this dependence we employ the method devised for matter clocks\cite{aps-prep}: 
we introduce the ''frozen'' time spaces: at each moment of time
the operator $\Theta_t$ is treated as time independent. Its spectral decomposition 
defines then the basis $\{e_{t,n}\}$ of the Hilbert space $\Hil_t$ of the initial data at time $t$.
The physical state is then expressed via positive/negative frequency spectral profiles 
$\tilde{\Psi}^{\pm}(t)$
\begin{equation}\label{eq:Psi-dec}
  \Psi(t,\phi) = \sum_{n=0}^{\infty}\left[ 
    \tilde{\Psi}^+_n(t) e_{t,n}(\phi) e^{i\omega_n(t)t}
    + \tilde{\Psi}^-_n(t) \bar{e}_{t,n}(\phi) e^{-i\omega_n(t)t} \right] ,
\end{equation}
further subject to \eqref{eq:constr}. The constraint itself translates into the set 
of countable number of coupled ordinary differential equations (ODEs) for 
$\tilde{\Psi}^{\pm}_n(t)$. The examination of $\Theta_t$ reveals the textbook result:
the spaces $\Hil_t$ correspond to a harmonic oscillator. The bases 
$e_{t,n}$ and their (time dependent) frequencies $\omega_n(t)$ are
\begin{equation}
  e_{t,n}(\phi) = N_{t,n} e^{-\frac{\alpha mv}{2}\phi^2} H_n(\sqrt{\alpha mv}\phi) ,
  \quad
  \omega_n(t) = ({12\pi G})^{-1/2} \sqrt{{\alpha mv}(2n+1)} , 
\end{equation}
where $H_n$ is the $n$th Hermite polynomial and $N_{t,n}$ are the normalization factors.

Each space $\Hil_t$ is unitary equivalent to $L^2(\re,\rd\phi)$ thus the 
physical inner product can be defined via selecting a time $t_o$ and setting 
$\langle\Psi|\Phi\rangle = \int \bar{\Psi}(t_o,\phi)\Phi(t_o,\phi) \rd\phi$. This 
inner product can be expressed as a product on $\Hil_t$ via 
$\langle\Psi|\Phi\rangle = \sum_{n=0}^{\infty} \sigma_n(t) \left[ %
\bar{\tilde{\Psi}}^+_n(t) \tilde{\Phi}^+_n(t) %
+ \bar{\tilde{\Psi}}^-_n(t) \tilde{\Phi}^-_n(t) \right]$
where the measure $\sigma_n(t)$ is fixed by the initial condition $\sigma_n(t_o)=1$
and the unitarity of the evolution. This in turn allows to easily 
construct the Dirac observables out of the kinematical ones.

Our main point of focus is the effect of the choice of the evolution parameter 
on the properties of the ground state. Since here the operator $\Theta_t$ is free from 
factor ordering ambiguities, this ground state is uniquely defined. To probe 
its gravitational effect we evaluate its energy density at given moment 
of time. It is
\begin{equation}
  \rho_o(t) = \langle \Psi(t,\cdot) | \hat{\rho} | \Psi(t,\cdot) \rangle
  = m [2V(t)]^{-1} > 0, \quad V(t) = a^3(t)
\end{equation}
thus its value is isolated from zero and scales as $a^{-3}$. Therefore the ground 
state of a single inflaton field  exerts the gravitational effect of the dust. 
For the models with infinite number of massive field modes this 
remnant is renormalized out via Fock space construction\cite{mgm-gowdy1}. 
However, it is believed that 
in LQC the volume parametrization would allow for finite number of modes only, 
rendering the vacuum energy non-removable and thus affecting (possibly significantly) 
the dynamics.

\noindent{{\bf Momentum deparametrization:}}  
The construction of the description using $b$ as the internal time is very similar 
to the one above, although now in order to avoid problems related with 
operator ordering we perform the deparametrization at the classical level, rewriting 
the Hamiltonian constraint \eqref{eq:constr-class} as the equation
\begin{equation}\label{eq:constr-b-class}
  v^2 = p_{\phi}^2 [3\pi Gb^2-\alpha^2m^2\phi^2]^{-1} .
\end{equation}
The subsequent Schr\"odinger quantization of the scalar field leads to the time dependent
equation of Klain-Gordon type
\begin{equation}\label{eq:constr-b}
  \partial^2_b \Psi(b,\phi) = -\Theta_b\Psi(b,\phi) , \quad 
  {\rm Dom}(\Theta_b)\subset \Hil_b \subset L^2(\re,\rd\phi) ,
\end{equation}
with $\Hil_b$ being the Hilbert space of the initial data at time $b$.
The operator $\Theta_b$ is (by inspection) essentially self-adjoint and the positive 
part of its spectrum is discrete. Therefore we can define the evolution as in $v$-time 
case, introducing the analog of the decomposition \eqref{eq:Psi-dec} and rewriting 
\eqref{eq:constr-b} as set of ODE's for spectral profiles. The construction of the 
physical inner product and the observables is also the same.

Although $\Theta_b$ is quite complicated, in frozen time formalism there exist the 
coordinate $x(b,\phi)$ such that it takes the form $\Theta_b = \frac{\alpha m}{12\pi Gb^2} %
\partial_x \sgn(|x|-\pi/4)\partial_x$, thus the basis elements $e_{b,n} \in %
\sgn(|x|-\pi/4)C^1(x)$. Given that, one can again calculate the gravitational effect of 
the ground state. Here it behaves like a massless scalar. Unlike in the $v$-time 
case however, the factor ordering freedom gives hope to bring the ground state energy
to zero.

\noindent{{\bf Application to LQC:}} As in its present form the $b$-time 
construction involves classical deparametrization, it is difficult to implement 
it directly in loop quantization. We remember however that the Hamiltonian constraint 
has to be regularized before quantization. We thus can implement a quasi-heuristic approach,
introducing a deparametrization after the regularization but still on the classical level. 
Then all the construction performed for geometrodynamics can be directly repeated
to completion. 
The only difference is the exact form to the time dependence of 
$\Theta_b$ as $b$ in \eqref{eq:constr-b-class} is now replaced with $\sin(b)$.

\bibliographystyle{ws-procs975x65}
\bibliography{main}

\end{document}